\documentclass[a4paper]{panl}
\usepackage{cite}
\usepackage{wrapfig}
\usepackage{graphicx}
\usepackage{amssymb}
\usepackage{amsfonts}
\usepackage{amsmath}
\usepackage{longtable}
\usepackage{rotating}
\usepackage{lscape}
\usepackage{epsfig}
\usepackage{multirow}
\def\bfg #1{{\mbox{\boldmath $#1$}}}

\originalTeX
\begin{document}

\title{Test of $T$-invariance in scattering of polarized protons
on tensor-polarized deuterons at energies of the NICA SPD\\
Проверка $T$-инвариантности в рассеянии поляризованных протонов
на тензорно-поляризованных дейтронах при энергиях NICA SPD
}
\maketitle
\authors{Yu.N.\,Uzikov$^{a,b,c}$\footnote{E-mail: uzikov@jinr.ru},
M.N.\,Platonova$^{d,a}$}
\setcounter{footnote}{0}
\from{$^{a}$\, V.P. Dzhelepov Laboratory, Joint Institute for Nucelar Researches, Dubna, 141980  Russia}
\from{$^{a}$\,Лаборатория ядерных проблем  им. В.П. Джелепова, Объединенный институт ядерных исследований, Дубна, 141980  Россия}
\from{$^{b}$\,M. V. Lomonosov
 Moscow State University, Faculty of Physics, 119991 Moscow, Russia}
\from{$^{b}$\, Физический факультет МГУ
им.~М.\,В.~Ломоносова,  Москва, 119991, Россия}
\from{$^{c}$\, Dubna State University, 141980 Dubna, Russia}
\from{$^{c}$\, Государственный университет ``Дубна'', 141980 Dubna, Россия}
\from{$^{d}$ \, Научно-исследовательский институт  ядерной физики имени Д. В. Скобельцына, МГУ
имени~М.\,В.~Ломоносова, Москва, 119991, Россия}
\from{$^{d}$ \,D. V. Skobeltsyn
 Institute of Nuclear Physics, M. V. Lomonosov Moscow State University, 119991 Moscow, Russia}

\begin{abstract}
Эффект нарушения $T$-инвариантности при условии сохранения $P$-четности дается полным сечением
взаимодействия векторно-поляризованной частицы с тензорно-поляризованной мишенью.
 Разработанный нами ранее формализм для расчета этого эффекта на основе спин-зависящей теории
Глаубера упругого $pd$-рассеяния
 использован здесь для расчета обсуждаемого эффекта
 в интервале энергий столкновения,  соответствующих  инвариантной массе
$pN$-системы  $\sqrt{s_{pN}}=5$--$30$ ГэВ. Необходимые для этого расчета спин-зависящие амплитуды упругого
 $pN$-рассеяния взяты из существующих
 феноменологических моделей $pN$-рассеяния
в рассматриваемой области энергий.

\vspace{0.2cm}

The effect of violation of $T$-invariance, provided that $P$-parity is preserved, is given by the total cross section
of the interaction of a vector-polarized particle with a tensor-polarized target.
A formalism for calculating this effect  developed previously and  based on the  spin-dependent Glauber theory
of elastic $pd$ scattering
 is used here to calculate the effect
under discussion in the range of collision energies corresponding to
the invariant mass of the $pN$ system $\sqrt{s_{pN}}=5$--$30$ GeV. The spin-dependent amplitudes of elastic $pN$ scattering
required for this calculation
 are taken from existing phenomenological models for $pN$ scattering
in the energy region considered.
\end{abstract}
\vspace*{6pt}


\label{sec:intro}
\section*{\bf Введение}

Барионная асимметрия Вселенной  $\eta$, определяемая как разность числа барионов и антибарионов, отнесенного к числу реликтовых фотонов в единице объема, имеет значение $\eta=6.7\times 10^{-10}$. Как показывает анализ \cite{Riotto:1999yt, Bernreuther:2002uj}, это значение примерно на 9 порядков величины превышает значение, ожидаемое из оценок по Стандартной космологической модели при использовании известной на сегодня  величины эффекта нарушения $CP$-инвариантности, обнаруженного в системе каонов, $B$- и $D$-мезонов. Это означает, что  в природе имеется не обнаруженный  пока  еще  дополнительный источник (источники) $CP$-нарушения.  При условии $CPT$-симметрии эти эффекты эквивалентны нарушению $T$-инвариантности.  Поиск таких  эффектов проводится  в экспериментах по  измерению постоянного электрического дипольного момента (ЭДМ)
нейтрона,  протона, легчайших ядер и нейтральных атомов, обнаружение которого
 свидетельсвовало бы о нарушении одновременно $T$- и  $P$-инвариантности.
 Другим аналогичным  источником может быть $CP$-нарушение в  лептонном секторе, которое исследуется в нейтринных экспериментах
 при измерении $CP$-нарушающей фазы  матрицы  смешивания  нейтрино. Экспериментальные ограничения на эти  эффекты систематически улучшаются \cite{Vergeles:2022mqu}.
 С другой стороны,  эффектам нарушения $T$-инвариантности, в которых сохраняется
 $P$-четность, Time-invariance Violating Parity Conserving (TVPC),
  и также сохраняется флэйвор, уделено значительно меньше внимания. Имеющиеся экспериментальные ограничения на TVPC  эффекты   более слабые,  чем на ЭДМ \cite{Huffman:1996ix,
  Vergeles:2022mqu}. Такие взаимодействия были предложены еще в 1965 году
\cite{Okun:1965tu} для объяснения  $CP$-нарушения  в физике каонов. Потенциально  это взаимодействие  может дать возможность для объяснения
наблюдаемой
барионной асимметрии.  В частности, сигналом TVPC
является полное сечение взаимодействия  поперечно поляризованного
($P_y$)
нуклона или ядра с
тензорно-поляризованным дейтроном ($P_{xz}$)
\cite{Barabanov:2004cr,Uzikov:2018tgq,Nikolaev:2020wsj}.
 Особый  интерес представляет  взаимодействие нуклонов  с ядрами дейтерия при энергиях, отвечающих ранней барионной Вселенной.
 В данной работе мы  приводим результаты расчета  энергетической зависимости эффекта TVPC  при  энергиях, отвечающих эксперименту  NICA SPD.

\section{\bf Элементы формализма}
\label{formuli}
Для описания упругого протон-протонного рассеяния необходимы пять независимых спиральных амплитуд:
\begin{eqnarray}
 \phi_1(s,t)=<++|M|++>, \nonumber \\
 \phi_2(s,t)=<++|M|-->, \nonumber \\
 \phi_3(s,t)=<+-|M|+->, \nonumber \\
 \phi_4(s,t)=<+-|M|-+>, \nonumber \\
 \phi_5(s,t)=<++|M|+->.
 \label{phi1-5}
\end{eqnarray}
Рассеяние нетождественных нуклонов  требует введения шестой амплитуды
$\phi_6$  с однократным переворотом спина, которая вырождается в $-\phi_5$ для тождественных нуклонов \cite{Buttimore:1978ry}.
  При этом связь с инвариантным  дифференциальным сечением
  \begin{equation}
   \frac{d\sigma}{dt}=K\{|\phi_1|^2+|\phi_2|^2 +|\phi_3|^2+|\phi_4|^2
   +4 |\phi_5|^2\}
   \label{dsdt}
  \end{equation}
 и векторной анализирующей способностью  $A_N$
 \begin{equation}
  A_N \frac{d\sigma}{dt}= -2 K {\rm Im}\{\phi_5^*(\phi_1+\phi_2+\phi_3-\phi_4)\}
  \label{AN}
 \end{equation}
  включает  нормировочный множитель $K$.
%
  Учитывая изоcпиновую структуру и соотношения, обусловленные $G$-четностью,
  вклады $\omega$, $\rho$, $f_2$, $a_2$  траекторий Редже и померонного обмена
  $P$ в амплитуды  упругого $pp$- и  $pn$-рассеяния  даются следующими соотношениями:
 \begin{eqnarray}
  \phi(pp)= -\phi_\omega -\phi_\rho + \phi_{f_2} +\phi_{a_2} +\phi_P, \nonumber \\
   \phi(pn)= -\phi_\omega +\phi_\rho + \phi_{f_2} -\phi_{a_2} +\phi_P.
\label{pn}
   \end{eqnarray}

  Спиральные амплитуды (\ref{phi1-5}) используются при формулировке моделей $pN$-рассеяния в области высоких энергий (см. \cite{Sibirtsev:2010cjv},
  \cite{Selyugin:2021cem}),
при этом нормировочный множитель $K$ в (\ref{dsdt})  в разных работах выбирается по-разному.
  При описании упругого $pd$- \cite{Platonova:2010wjt} и $^3$He$d$-рассеяния \cite{Platonova:2023epi} в спин-зависящей теории Глаубера
aдронные $T$-четные $P$-четные спиновые амплитуды $pN$-рассеяния были использованы в форме
\begin{eqnarray}
M_N=A_N+C_N{\bfg \sigma}_p\cdot {\bf \hat n}+C'_N{\bfg \sigma}_N \cdot {\bf \hat n},\nonumber \\
 + B_N({\bfg \sigma}_p\cdot {\bf \hat  k})({\bfg \sigma}_N\cdot {\bf \hat  k}),\nonumber \\
 +(G_N+H_N)(\bfg \sigma_p\cdot {\bf \hat q})(\bfg \sigma_N\cdot {\bf \hat q}),\nonumber\\
 +(G_N-H_N)(\bfg \sigma_p\cdot {\bf \hat  n})(\bfg \sigma_N\cdot {\bf \hat n})
 \label{MpN};
 \end{eqnarray}
  здесь  ${\bfg \sigma}_p$ (${\bfg \sigma}_N)$ --  спиновые матрицы Паули, действуюшщие  на спиновое состояние протона пучка (нуклона мишени $N$), а единичные орты ${\bf \hat  k}, {\bf \hat  q}$ и  ${\bf \hat  n}$ определены через начальный  ${\bf p}$ и конечный  ${\bf p'}$  импульсы рассеивающегося  протона:
 $\bf \hat k= {(\bf p+\bf p')}/{|\bf p+\bf p'|},\bf \hat q =  {(\bf p-\bf p')}/{|\bf p-\bf p'|},
  \bf \hat n=[ \bf \hat k\times  \bf \hat q]$.
  Спиновые амплитуды $A_N$, $B_N$, $C_N$, $G_N$  и $H_N$  связаны  со спиральными амплитудами $\phi_1\div \phi_5$ следующими соотношениями
  \cite{Platonova:2010wjt}:
  \begin{eqnarray}
A_N=(\phi_1+\phi_3)/2, \ \
B_N=(\phi_3-\phi_1)/2, \nonumber \\
C_N=i\phi_5, \ \
G_N=\phi_2/2, \ \
G_N=\phi_4/2, \nonumber \\
C'_N=C_N+i\frac{q}{2m_N}A_N.
\label{A-phi}
\end{eqnarray}
Последнее соотношение в (\ref{A-phi})  выполняется    при малых углах рассеяния в области высоких энергий (см. работу \cite{Platonova:2010wjt} и ссылки в ней).

 В теории Глаубера вклад в амплитуду $pA$-рассеяния  вносят только $pN$-амплитуды на массовой поверхности.
Мы  рассматриваем здесь следующие три члена $t$-матрицы TVPC упругого $pN$-рассеяния,
не исчезающие на массовой поверхности \cite{Beyer93}:
 \begin{eqnarray}
t_{N}=
 h_{N}[({\bfg \sigma}_p\cdot {\bf \hat k})(\bfg \sigma_N\cdot {\bf \hat q})+
 ({\bfg \sigma}_p\cdot {\bf \hat q})({\bfg \sigma}_N\cdot {\bf \hat k})
-\frac{2}{3}({\bfg \sigma}_N\cdot {\bfg \sigma}_p)({\bf q} \cdot {\bf \hat  k)}] \nonumber \\
+g_{N}[{\bfg \sigma}_p\times{\bfg\sigma}_N]\cdot [{\bf\hat  q}\times {\bf \hat k}]({\bfg \tau}_p - {\bfg \tau}_N)_z \nonumber \\
+g_{N}'({\bfg \sigma}_p-{\bfg \sigma}_N)\cdot i[{\bf \hat q}\times{\bf\hat  k}]
[{\bfg \tau}_p\times{\bfg \tau}_N]_z;
\label{tpN}
\end{eqnarray}
 здесь $h_{N}$, $g_{N}$, $g_{N}'$ --  амплитуды, содержащие  неизвестые константы TVPC $NN$-взаимодействия, а
 ${\bfg \tau}_p$ (${\bfg \tau}_N$) -- изоспиновые матрицы Паули, действующие на  состояние  начального протона
  (нуклона $N$).

 TVPC эффект, в
англоязычной
  литературе называемый нуль-тест сигналом,
 в $pd$- ($^3$He$d$-) рассеянии
 определяется мнимой частью TVPC амплитуды $pd$- ($^3$He$d$-) рассеяния на нулевой угол $\tilde g$
  \cite{Uzikov:2015aua}:
 \begin{equation}
 \sigma_{TVPC} = -4\sqrt{\pi}\frac{2}{3} {\rm Im}\tilde g.
 \label{sigmatvpc}
 \end{equation}
Амплитуда $\tilde g$ вычисляется в теории Глаубера
и, согласно работе \cite{Uzikov:2016lsc}, для процесса $pd \to pd$
  имеет следующий вид:
 \begin{eqnarray}
  {\tilde g} = \frac{i}{4\pi m_p}\int_0^\infty dq q^2[ S_0^{(0)}(q)
  -\sqrt{8}S_2^{(1)}-4S_0^{(2)}(q)
  +9S_1^{(2)}(q)\nonumber \\
  +\sqrt{2}\frac{4}{3}S_2^{(2)}(q)]
  [-C'_n(q)h_p+C'_p(q)(g_n-h_n)],
  \label{gtilde}
 \end{eqnarray}
  где
  $C^\prime_N(q)$ --
 амплитуда  $pN$-рассеяния с однократным переворотом спина,  определенная  в  выражении (\ref{MpN}),
$h_N$ ($g_N$) --  TVPC амплитуды $pN$-рассеяния из \eqref{tpN},
$S_n^{(m)}(q)$ $(m,n=0,1,2)$ --  формфактор дейтрона,  определенный в
\cite{Uzikov:2016lsc}
с учетом вклада $S$- и $D$-
компонент дейтронной волновой функции.
Cоответствующая TVPC амплитуда упругого $^3$He$d$-рассеяния
  также  определяется  механизмом
  двукратного рассеяния \cite{Uzikov:2023eex} и имеет аналогичный вид,
  с тем отличием от $pd$, что вместо $pN$-амплитуд
в выражение для $\tilde g$
  входят амплитуды $^3$He$N$-рассеяния, которые были получены в работе
  \cite{Platonova:2023epi}.

\label{results}
\section{\bf Численные результаты}

  В численных расчетах
мы используем две
феноменологические модели для спин-зависящих амплитуд  упругого $pN$-рассеяния. Одна из них дается реджевской параметризацией $pp$-данных о дифференциальном  сечении и спиновых наблюдаемых $A_N$, $A_{NN}$.
   Эта модель включает вклады  четырех  траекторий Редже, $\omega$, $\rho$, $f_2$, $a_2$, и померонный обмен $P$
   \cite{Sibirtsev:2010cjv}. Как отмечено в предыдущем разделе, в Редже-модели амплитуды $pp$- и $pn$-рассеяния даются  разными линейными
   комбинациями этих пяти вкладов (\ref{pn}). Поэтому для построения $pn$-амплитуд
   на основе  результатов полученной  в работе \cite{Sibirtsev:2010cjv} параметризации
   для $pp$-амплитуд мы используем второе соотношение в (\ref{pn}).
   Область применения  параметризации \cite{Sibirtsev:2010cjv} ограничена
  интервалом  значений импульса  протона в лабораторной системе отсчета
  $p_l= 3\div  50$ ГэВ/$c$,  что соответствует интервалу инвариантной массы $pp$-системы $\sqrt{s_{pp}}= 2.8 \div 10$ ГэВ.

   Во втором используемом здесь подходе
-- HEGS (High Energy Generalized Structure,
см. работу \cite{Selyugin:2021cem} и ссылки в ней)
yпругое рассеяние
 $pp$, $p\bar{p}$ и $pn$ на малые углы
    рассматривается
   в развиваемой автором \cite{Selyugin:2021cem}
   Редже-эйкональной модели, при этом  структура
   нуклонов учитывается путем включения данных об обобщенных партонных распределениях нуклонов.
   Полученные в этой модели спиральные амплитуды упругого нуклон-нуклонного рассеяния
   позволяют   описать  имеющиеся
   экспериментальные данные по дифференциальному сечению и односпиновой асимметрии $pp$-рассеяния
   $A_N(s,t)$ в интервале энергий $\sqrt{s}$ от 3.6 до 10
   ТэВ с минимумом варьируемых параметров.

  В обеих моделях при рассматриваемых здесь энергиях $\sqrt{s_{pp}}\ge 3$ ГэВ
для спиральных амплитуд  упругого $pp$-рассеяния выполняются следующие
соотношения: $\phi_1=\phi_3$, $\phi_2=0,\phi_4=0$.

При расчете TVPC-сигнала  по оптической теореме
в используемых $pp$-амплитудах исключены кулоновские вклады. Дело в том, что кулоновское взаимодействие не нарушает $T$-инвариантности и поэтому не может дать прямой вклад в TVPC сигнал \cite{Uzikov:2015aua}.

\begin{figure}[t]
\begin{center}
\includegraphics[width=127mm]{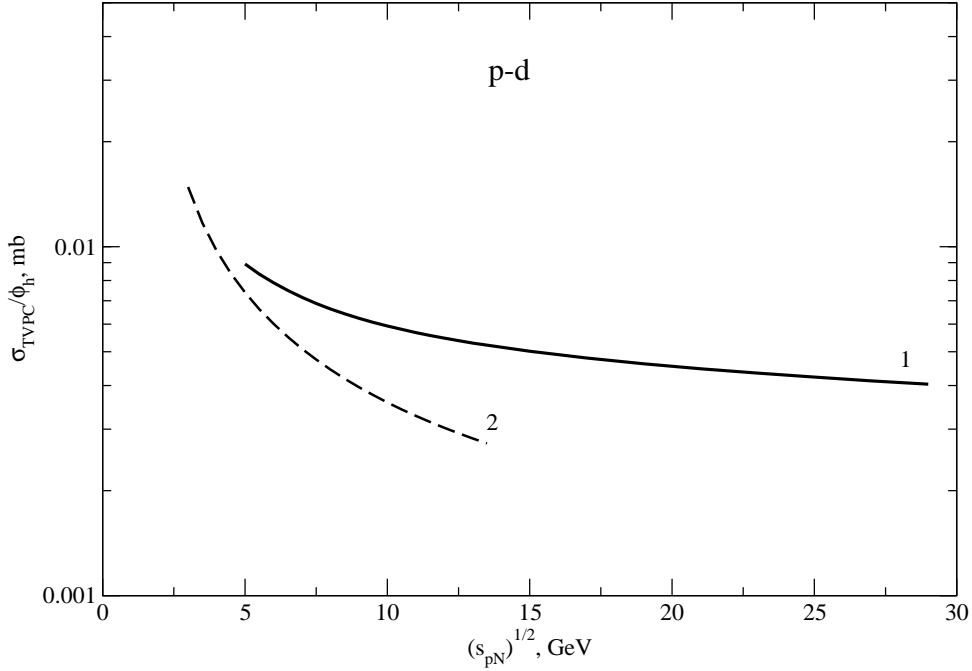}
\vspace{-3mm}
\caption{ Зависимость от энергии  TVPC-сигнала в  $pd$-рассеянии
$\sigma_{TVPC}$,
деленного на отношение $\phi_h$  констант
связи $h_1$-мезона с нуклоном для TVPC и обычного $T$-четного $P$-четного  взаимодействий.
 Использованы  две  разные модели спиновых $pN$-амплитуд: 1 -- HEGS \cite{Selyugin:2021cem},  2 -- реджеонная параметризация \cite{Sibirtsev:2010cjv}.
По оси абцисс отложена  инвариантная  масса взаимодействующей  $pN$-пары -- налетающего протона и нуклона мишени $N$.
}
\end{center}
\labelf{fig01}
\vspace{-5mm}
\end{figure}
Как показано в работах \cite{Uzikov:2015aua, Uzikov:2016bja, Uzikov:2017dns}, взаимодействие типа $g'_{N}$ не дает вклада в TVPC сигнал в $pd$-рассеянии,
а TVPC сигналы для  взаимодействий  $h_{N}$- и $g_{N}$-типов  имеют сходное поведение в зависимости от энергии.
Следуя работам \cite{Beyer93,Uzikov:2015aua}, для амплитуды
 $h_{N}$-типа, связанной с обменом аксиальным  мезоном  $h_1(1170)$,
 взято следующее выражение:
 \begin{equation}
  h_N = -i\phi_h\frac{2G_h^2}{m_h^2+{\bf q}^2} F_{hNN}({\bf q}^2),
 \end{equation}
где $\phi_h=\tilde G_h/G_h$ -- отношение константы связи $h_1$-мезона
с нуклоном для $T$-неинвариантного взаимодействия ($\tilde G_h$) к соответствующей константе $T$-инвариантного взаимодействия ($G_h$);
$F_{hNN}({\bf q}^2)=(\Lambda^2-m_h^2)/(\Lambda^2-{\bf q}^2)$ --
феноменологический  монопольный формфактор  в  вершине $hNN$. Численные параметры взяты из
\cite{Beyer93}:
$m_h=1.17$ ГэВ, $\Lambda=1$ ГэВ, $G_h=4\pi\times 1.56$.
Результаты расчетов TVPC сигнала для взаимодействия $h_{N}$-типа
представлены на рисунке для двух
моделей $pN$-рассеяния \cite{Sibirtsev:2010cjv}
 и \cite{Selyugin:2021cem}
в зависимости от инвариантной массы
пары взаимодействующих нуклонов -- налетающего протона и нуклона $N$ в дейтроне.
Полученные результаты показывают, что с ростом энергии  величина сигнала убывает,
при этом скорость убывания  существенно зависит от используемой модели адронного $T$-инвариантного $P$-инвариантного $pN$-взаимодействия.
Следует отметить, что модельно-независимые  амплитуды  упругого $pN$-рассеяния,
 полученные в результате фазового  анализа  соответствующих поляризационных экспериментальных данных, имеются в базе данных SAID \cite{Arndt:2007qn}. Максимальная кинетическая энергия для $pn$-рассеяния в этой базе составляет 1.2 ГэВ, что соответствует $\sqrt{s_{pn}}=2.4$ ГэВ.
Выполненный нами c использованием  базы данных SAID расчет TVPC сигнала в $pd$-рассеянии при этой энергии
дает значение
в $\sim$ 7 раз выше,
чем его  максимальное значение, приведенное на рисунке
при $\sqrt{s_{pp}}=3$ ГэВ для параметризации  \cite{Sibirtsev:2010cjv}. Это убываниие TVPC эффекта соответствует общей тенденции ослабления спиновых эффектов  в адронных взаимодействиях с ростом энергии.

\label{sec:conclusion}
\section*{\bf Заключение}

Сигнал нарушения $T$-инвариантности при сохранении $P$-четности (TVPC) в рассеянии поляризованного фермиона на тензорно-поляризованной мишени определяется неизвестной
константой взаимодействия, не входящего в Стандартную модель. Ожидается, что это взаимодействие осуществляется на очень малых
межнуклонных расстояниях
и поэтому его интенсивность слабо зависит от энергии.
Однако TVPC константа под знаком интеграла
в  выражении для искомого сигнала (\ref{gtilde}) домножается на фактор, определяемый обычными $T$- и $P$-инвариантными адронными взаимодействиями,
спиновые эффекты в которых довольно сильно зависят от энергии, убывая с ее ростом.
Выполненные в данной работе расчеты в интервале энергий
$\sqrt{s_{pp}}=3 \div 25$ ГэВ, соответствующих  области  готовящегося в Дубне эксперимента по спиновой физике NICA SPD,  показывают, что степень  уменьшения TVPC сигнала в существенной  мере  определяется  используемой моделью для адронных спиновых амплитуд $pN$-рассеяния.
\newpage
\label{sec:acknowledgement}
\section*{\bf Благодарности}

Мы благодарны О.В. Селюгину за предоставленные им численные значения  спиновых $pN$-амплитуд, полученные в его модели.

\label{sec:funding}
\section*{\bf Финансовая поддержка}
Исследование выполнено за счет средств Российского Научного Фонда, грант № 23-22-00123, https://rscf.ru/en/project/23-22-00123/.

\label{sec:conflict}
\section*{\bf Конфликт интересов}
Авторы этой работы утверждают, что у них нет конфликта интересов.

\bibliographystyle{pepan}
\bibliography{Uz-Pl-Sess-m3.bib}
\end{document}